\documentclass[fleqn,12pt,twoside]{article}
\usepackage[headings]{espcrc1}
\usepackage{natbib}

\readRCS
$Id: espcrc1.tex,v 1.2 2004/02/24 11:22:11 spepping Exp $
\usepackage{graphicx}
\usepackage[figuresright]{rotating}

\newcommand{\AmS}{{\protect\the\textfont2
  A\kern-.1667em\lower.5ex\hbox{M}\kern-.125emS}}

\title{Alfv\'en waves as a driving mechanism in stellar winds}

\author{A. A. Vidotto \address[IAG]{Instituto de Astronomia, Geof\'isica e Ci\^encias
Atmosf\'ericas - Universidade de S\~ao Paulo. \\ Rua do Mat\~ao,
1226, Cidade Universit\'aria, 05508-900, S\~ao Paulo - SP - Brazil.} and
               V. Jatenco-Pereira\addressmark}

\begin{document}

\maketitle

\begin{abstract}

Alfv\'en waves have been invoked as an important mechanism of particle acceleration in stellar winds of cool stars.
After their identification in the solar wind they started to be studied in winds of stars located in different regions 
of the HR diagram. We discuss here some characteristics of these waves and we present a direct 
application in the acceleration of late-type stellar winds.

\end{abstract}

\section{Introduction}

In 1942 combining fluid hydrodynamics and electromagnetism Hannes Alfv\'en discovered
a new mode of waves, thus opening a new field of research. He studied
the mutual interaction between conducting fluid motion and electromagnetic fields and discovered
these waves (Alfv\'en, 1942) that later on were named {\it{Alfv\'en waves}}. However, to be verified the 
waves needed to be detected in laboratory experiments. This was done in the late 40s 
and beginning of the 50s, in experiments with mercury and liquid sodium. It was discovered
that in liquid metals the waves are strongly damped. In the 60s, experiments in plasma demonstrated 
weakly damped Alfv\'en waves with the properties predicted by the theory. Since then this field
has expanded mainly due to space studies and thermonuclear research, making possible, among other things, 
to create high temperature plasma in laboratories. 

An Alfv\'en wave propagating in a plasma is a traveling oscillation of the ions and the magnetic field. The ion mass 
density provides the inertia and the magnetic field line tension provides the restoring force. The wave vector can either propagate in the parallel direction of the magnetic field or at oblique incidence. The waves efficiently carry energy and momentum along the magnetic field lines. 

Identification of Alfv\'en waves in the solar wind by means of spacecraft measurements was first achieved in late 60s.
In 2007, Tomczyk et al. (2007) reported the detection of Alfv\'en waves in 
images of the solar corona with the Coronal Multi-Channel Polarimeter instrument at the 
National Solar Observatory, New Mexico. Also in 2007, several research groups reconfirmed that Alfv\'en waves 
have sufficient energy
to heat the solar corona and momentum to accelerate the solar wind (De Pontieu et al., 2007; Okamoto et al., 2007). 
The Sun is a benchmark for stellar research providing an example 
of a cool main sequence star losing mass. 

Since the discovery of the high velocity winds from O and B supergiant stars by Morton (1967) 
and his co-workers (Morton, Jenkins \& Brooks, 1969), the theories of stellar winds had their interest 
amplified. On another part of the HR diagram, the winds from evolved K and M stars, characterized by low
velocities, demanded a different theoretical approach. Although presenting very different terminal velocities, 
in the range of 600 - 3500 km s$^{-1}$ for early-type and 10 - 100 km s$^{-1}$
for late-type K and M stars, these stars present a similarity: both luminous early and late-type stars are losing mass
at a rate as high as 10$^{-5}$ $M_\odot$ yr$^{-1}$ (Lamers \& Cassinelli, 1999). 

The idea that a flux of Alfv\'en waves could accelerate winds of stars in many regions
of the HR diagram remount from a few decades ago (Hollweg, 1973, 1987; Hartmann \& MacGregor, 1980;
Holzer, Fla \& Leer, 1983; Underhill 1983; Hartmann \& Avrett, 1984;
Jatenco-Pereira and Opher, 1989a,b,c; dos Santos, Jatenco-Pereira and Opher, 1993; 
Jatenco-Pereira, Opher and Yamamoto, 1994; among others). The belief that Alfv{\'e}n waves are present in 
the magnetized winds of cool
giant stars is supported by the fact that these waves are
observed in the solar wind (Smith et al., 1995; Balogh et al., 1995).
If there are oscillations in the magnetic field
at the base of the wind, Alfv{\'e}n waves will be generated. As they
propagate outward, the dissipation of their energy and the
transfer of their momentum to the plasma can heat and accelerate the wind. As an example of how these waves can be
used as an acceleration mechanism of a stellar wind, we show later on this paper a simple
model to explain the mass-loss in a typical supergiant K5 star (Section 3). In Section 2, 
we present the issue of the damping of the
Alfv\'en waves. In Section 4, we apply this model to a typical K5 supergiant star and in Section 5, we draw our
conclusions. 

\section{The damping mechanism for Alfv\'en waves}

For cool red giant stars the flux of Alfv\'en waves can provide an efficient driving mechanism
for the wind. In this mechanism it is the wave magnetic pressure that accelerates the wind. 
Hartmann \& MacGregor (1980) showed that if the waves are not damped, the resultant terminal velocities are 
much higher than observed. Assuming 
that the waves are damped with a constant damping length, they obtained terminal velocities 
and mass-loss rate consistent with the observations. 

In order to improve the model presented by Hartmann \& MacGregor (1980), 
Jatenco-Pereira \& Opher (1989a) discarded the use of an artificial constant
damping length for the waves and considered three different physical 
damping mechanisms: (i) the nonlinear damping, (ii) the resonant 
surface damping, and (iii) the turbulent damping. 
They showed that, for an isothermal atmosphere with $T \sim 10^4$~K, their
model can reproduce the observed large mass-loss rates and the small
ratios between the terminal velocity and the escape velocity at the base ($u_\infty/v_{e_{0}}$) 
of these stellar winds.

To illustrate this mechanism, we present a simplified version of this model (see Vidotto \& Jatenco-Pereira, 2006; Falceta-Gon\c calves, Vidotto \& Jatenco-Pereira, 2006 for a more complete analysis). We focus our study on the resonant surface damping for the Alfv{\'e}n waves. A surface wave exists at the interface separating two environments, e.g., open magnetic flux tubes. Magnetohydrodynamic surface waves may decay through a process called ``resonance absorption" (Hollweg, 1987), that leads to a  concentration of the surface wave energy into the thin resonant layer. In this layer, the energy may be dissipated locally as heat or concentrated in the form of turbulent Alfv{\'e}n waves which may propagate further and be dissipated elsewhere, or concentrated in other forms, such as high energy particles, electric currents, etc. In the present paper we suggest that resonant absorption dissipates the energy locally as heat with the damping length $L$ given by (Jatenco-Pereira \& Opher, 1989a)
\begin{equation} 
L = L_0 \left( \frac{v_A}{v_{A_{0}}} \right)^2
\sqrt{\frac{A(r_0)}{A(r)}} (1+M) \, , 
\end{equation}
where $L_0$ is the damping length of the wave at the base of the wind, $M=u/v_A$ is the 
Alfv{\'e}n-Mach number, the Alfv{\'e}n speed is given by $v_A = (B / \sqrt{4 \pi \rho})$ and $A(r)$ is the cross-section 
of a flux tube at a radial distance $r$. The index ``$0$'' indicates that the parameter is being evaluated at
the stellar surface, i.e., at $r = r_0$. If a large damping length is considered 
(i.e., there is a smooth damping of the wave throughout the wind), there will be a non negligible outward
force acting beyond the critical point of the wind, creating thus a high-speed wind. Hence, the damping length cannot 
be too large.
This will also assure that the WKB approximation is valid in the wind we study.

\section{Model}\label{sec.model}

The first attempts in accelerating winds by means of Alfv\'en waves considered that the waves were 
propagating in an isothermal ambient. This assumption simplifies the equations, since the energy equation
simply reads $T=$ constant. However, if we are interested in modeling the temperature profile of the wind,
there is a need to solve a consistent energy equation simultaneously with the 
mass and momentum equations of the wind. In this way,
we are able to obtain both the temperature and the velocity profiles
of the wind and compare them with observations. 

\subsection{Geometry of the Magnetic Field Lines}

\begin{figure}[t]
   \centering
   \includegraphics[scale=.5]{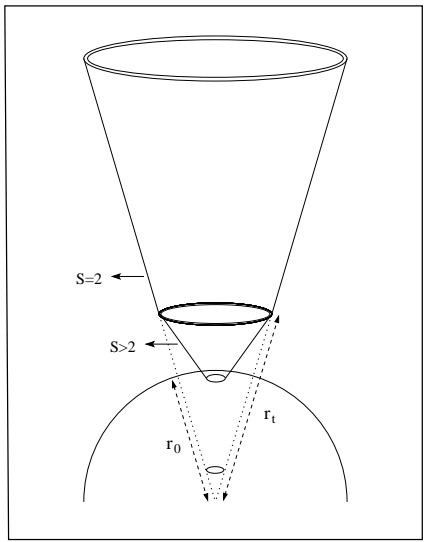}
      \caption{Magnetic field geometry used in the model (not in
  scale). The stellar radius is denoted by $r_0$. The magnetic field
  geometry is divergent up to the transition radius $r_t$, i.e., it
  has a superradial index $S > 2$. Beyond $r_t$, the magnetic field
  lines become radial and the index equals $2$.}
         \label{geometry}
   \end{figure}

For a long time, it is known that the coronal holes are the source of the high-speed solar
wind streams at the Earth's orbit (e.g., Peter \& Judge, 1999). 
Solar coronal holes observations show a superradial expansion of the magnetic flux tubes:
the area expansion of a solar coronal hole is $5-13$ times
greater than for a radial expansion (Lie-Svendsen, Hansteen \& Leer, 2002;
Esser et al., 2005; Tu et al., 2005).
Jatenco-Pereira \& Opher (1989a) implemented the Alfv\'en wave driven wind model by 
taking into account our knowledge on coronal holes in the Sun. We apply the same
assumption for cool giant stars. For simplicity, we assume 
that in the wind of a typical K5 supergiant star, the non-radial
expansion factor is  
\begin{equation}
F = \frac{\Omega}{\Omega_0} = 10 \, ,
\end{equation}
where $\Omega_0$ and $\Omega$ are the solid angles measured at the stellar surface
and at the edge of the coronal hole, respectively. Hence, in order to model the 
diverging geometry for the magnetic field, we follow the idea given by Kuin and Hearn (1982). 
Thus, the cross-section of a flux
tube at a radial distance $r$ is given by
\begin{displaymath}
A(r) = \left\{ \begin{array}{ll}
A(r_0) (r/r_0)^S & {\rm ~if~ } r \leq r_t \\
A(r_0) (r_t/r_0)^S (r/r_t)^2 & {\rm ~if~ } r > r_t \, ,
\end{array} \right.
\end{displaymath}
where $S$, the superradial index, is a parameter that determines the
divergence of the geometry from the stellar radius $r_0$ up to the
transition radius $r_t$, as shown in Figure~\ref{geometry}. For a
given $S>2$, $r_t$ is obtained by 
\begin{equation}
F = \frac{\Omega}{\Omega_0} = \frac{A(r_t)/r_t^2}{A(r_0)/r_0^2} =
\left( \frac{r_t}{r_0} \right)^{S-2} \, . 
\end{equation}
Hence, we have
\begin{equation}
r_t = F^{1/(S-2)} \, r_0 \, .
\end{equation}
Considering this geometry, the conservation of magnetic flux yields
a magnetic field intensity of
\begin{equation}
B(r) = B_0 \frac{A(r_0)}{A(r)} \, .
\end{equation}
A detailed study of the diverging geometry of the magnetic field in the winds of late-type stars can be found in Jatenco-Pereira \& Opher (1989a).

\subsection{Wind Equations}

The equation of mass continuity 
expresses the conservation of mass. If the flow velocity is denoted by $u$ and
the gas density is $\rho$, we have in steady-state 
\begin{equation}\label{mass}
\rho u A(r) = \mathcal{C} \, .
\end{equation}
The constant $\mathcal{C}$ in equation~(\ref{mass}) is computed at
the wind base, i.e., $\mathcal{C} = \rho_0 u_0 A(r_0)$.

Assuming a steady flow, the
equation of motion is written as
\begin{equation}\label{momentum}
\rho u \frac{du}{dr} = - \rho \frac{G M_\star}{r^2} - 
\frac{dP}{dr} - \frac{d}{dr}  \frac{\langle (\delta
  B)^2 \rangle}{8 \pi} \, ,
\end{equation}
where  $-\rho G M_\star / r^2$ is the gravitational attraction force, $-dP/dr$ is the pressure gradient, and $d[{\langle (\delta B)^2 \rangle}/(8 \pi)]dr$ is the Alfv{\'e}n wave magnetic pressure gradient. The parameter $\delta B$ is the magnetic field amplitude of the wave. It is
related to the energy density $\epsilon$ of the wave by 
\begin{equation}
\epsilon = \frac{\langle (\delta B)^2 \rangle}{4 \pi} \, .
\end{equation}

The energy equation is determined through the balance between wave
heating, adiabatic expansion, and radiative cooling
(Hartmann, Edwards, \& Avrett, 1982). Assuming an ideal gas, we can write the
gas pressure as $P = \rho k_B T / m$, where $k_B$ is the Boltzmann
constant, $T$ is the gas temperature, and $m$ is the mean mass per
particle. Thus, neglecting conduction, we write the energy equation as 

\begin{equation} \label{energy}
\rho u \frac{d}{dr} \left( \frac{u^2}{2} + \frac52 \frac{k_B T}{m} -
\frac{G M_\star}{r}  \right) + \frac{u}{2} \frac{d \epsilon}{dr} = (Q
- P_R) \, .
\end{equation}
The term $(u/2) d \epsilon/dr$ is the rate at which the waves do work
on the gas. $Q$ is the wave heating rate, i.e., the rate at which the
gas is being heated due to dissipation of wave energy, and $P_R$ is
the radiative cooling rate, both in erg~cm$^{-3}$~s$^{-1}$. The wave
heating can be written as (Hollweg, 1973)
\begin{equation}
Q = \frac{\epsilon}{L} (u + v_A)
\end{equation}
and the radiative cooling is given by
\begin{equation}
P_R = \Lambda \, n_e \, n_H \, ,
\end{equation}
where $n_e$ is the electron density, $n_H$ is the hydrogen density and
$\Lambda$ is the radiative loss function. Here, we adopt $\Lambda$
given by Schmutzler \& Tscharnuter (1993) and calculate $n_e$ as
Hartmann \& MacGregor (1980).

If the waves are damped, the wave energy is dissipated and we write
\begin{equation}
\epsilon = \epsilon_0 \frac{M_0}{M} \left( \frac{1+M_0}{1+M} \right)^2
\exp \left[ - \int_{r_0}^{r} \frac{1}{L} \, dr' \right] \, . 
\end{equation}
The initial energy density $\epsilon_0$ and the initial wave
flux $\phi_{A_{0}}$ are related to each other by (Jatenco-Pereira \& Opher, 1989a)
\begin{equation}
\phi_{A_{0}}  = \epsilon_0  v_{A_{0}} \left( 1 + \frac32 M_{0} \right) \, .
\end{equation}

From the equations above, we write the temperature variation as
\begin{equation} \label{energy2}
\frac{dT}{dr} = \frac23 \frac{T}{r} \left[ \frac{r(Q-P_R)}{\rho u (k_B
    T /m)} - \left( Z + \frac{r}{u} \frac{du}{dr} \right) \right] \, , 
\end{equation}
where, we defined $Z$ as
\begin{displaymath}
Z = \left\{ \begin{array}{ll}
S & {\rm ~if~} r \leq r_t \\
2 & {\rm ~if~} r > r_t \, .
\end{array} \right.
\end{displaymath}
As $\epsilon = \rho \langle (\delta v)^2 \rangle$,
where $\delta v$ is the velocity fluctuation of the wave, the
velocity gradient is given by
\begin{eqnarray}\label{momentum2}
\frac{1}{u} \frac{du}{dr} \left[ u^2 - \frac53 \frac{k_B T}{m} -
 \frac{\langle (\delta v)^2 \rangle}{4} \left( \frac{1+3M}{1+M}
 \right)   \right] = \frac{Z}{r} \left[ \frac53
 \frac{k_B T}{m} - \frac23 \frac{r(Q-P_R)}{Z \rho u} -  \frac{G
 M_\star}{rZ} + \right. \nonumber \\ \left. + \frac{\langle (\delta
 v)^2 \rangle}{2LZ} r + \frac{\langle (\delta v)^2 \rangle}{4} \left(
 \frac{1+3M}{1+M} \right) \right]\, .
\end{eqnarray}

Thus, in order to obtain self-consistently the velocity and the temperature 
profiles of the wind, it is necessary to solve equations~(\ref{mass}),
(\ref{energy2}), and (\ref{momentum2}) simultaneously. 

\section{Results and discussions}\label{sec.results}

The model is applied here to the wind of a
typical K5 supergiant star with a mass of 16~M$_\odot$ and a radius of 400~R$_\odot$. 
We adopt a temperature of $3500$~K at the base of the wind, and assume that the magnetic
field intensity is $\sim 10$~G at this location. In our model, we adopt an initial wave flux of $\phi_{A_{0}} \sim
10^6$~erg~cm$^{-2}$~s$^{-1}$, which is of the same order of magnitude
as the one estimated for the Sun (Banerjee et al., 1998), and take $L_0 = 0.2~r_0$ and $S = 5.2$.
The initial density is $\rho_0 = 1.07 \times 10^{-13} {\rm g~cm}^{-3}$. 

We perform the calculations until $r=300 \, r_0$, where the
terminal velocity has been already reached.  

\subsection{Typical K5 supergiant star}

Similarly to the wind first studied by Parker (1960), the 
momentum equation (eq.~[\ref{momentum2}]) has a
critical point when the term in brackets on the left-hand side becomes zero, and hence the bracketed term on the right-hand side must necessarily be zero too. The
requirement that the wind velocity should increase through the
critical point determines the initial velocity at the base of the
wind, i.e., ${du}(r=r_0)/{dr} \geq 0$. Hence,
to find the critical solution for a given set of initial conditions,
we iterate the initial velocity until the solution passes through the
critical point. Together with the initial density, the initial
velocity determines the mass-loss rate.

Our results are shown in Figure 2, where the velocity and
the temperature profiles obtained were plotted for $r \leq 30~r_0$. We
obtained a terminal 
velocity of $u_\infty \simeq 63$~km~s$^{-1} \simeq 0.5 v_{e_{0}}$ and
a mass-loss rate of $\dot{M} \simeq 1.2 \times
10^{-7}$~M$_\odot$~yr$^{-1}$. These results compares favorably with
observations: observationally, the terminal velocity should be lower than the
surface escape velocity in a ratio around 1:2, and the mass-loss rate 
should range from $10^{-8}$ to $10^{-5}$ M$_\odot$ yr$^{-1}$ (Linsky 1998).  

   \begin{figure}[t]
   \centering
   \includegraphics[scale=.4]{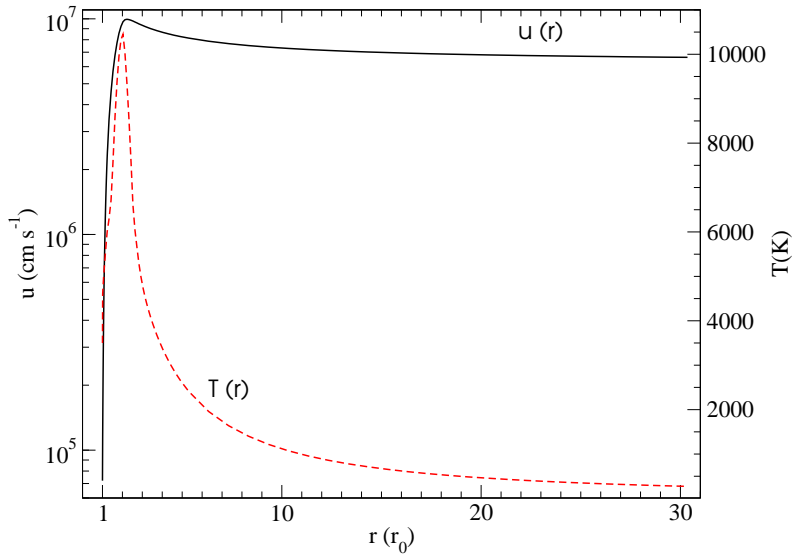}
      \caption{Velocity (solid line) and temperature (dashed line) profiles
  obtained for the wind of a 
  typical K5 supergiant star.}
         \label{vel_temp}
   \end{figure}

According to Roddier \& Roddier (1985), the CII emission detected in cool winds 
of giant stars is associated to regions with temperatures of $7000-9000$~K. 
Carpenter, Robinson, \& Judge (1995)
obtained, for the wind of $\gamma$~Cru (M3.4III), a maximum temperature
of $\sim 9000$~K. Carpenter et al. (1999) found a maximum temperature of $\sim
10000$~K for the wind of $\lambda$~Vel (K5Ib). For $\zeta$~Aur (K4Ib),
Eaton (1993) found that the wind temperature reaches $12000$~K at $\sim
2 \, r_0$. We can see in Figure 2 an abrupt rise in the temperature of the wind: the
temperature rises from $3.5 \times 10^3$~K to $\sim 10^4$~K in the
range from $1 ~ r_0$ to $\sim 2 ~ r_0$. This heating is mainly due to
the dissipation of the Alfv{\'e}n waves, ie, caused by the wave damping. At large
distances, when almost all the flux of Alfv\'en waves has been dissipated, 
the flow expands adiabatically. Hence, in the absence of
strong heating, the temperature tends to fall monotonically with the
adiabatic exponent $4/3$.

As to exemplify the effects of the choice of $L_0$ in our model, we have run a second model taking into account a different value of initial damping length: $L_0=0.4~r_0$. Comparing this second model with the previous one, we obtained a value of $\sim 100$ km/s while for the first model a flow with a terminal velocity of $\sim 63$ km/s was achieved. This situation is explained by the position where a large fraction of the energy is deposited in the wind. If energy is mainly deposited in the super-critical part of the flow, as is the case for the second model, the initial structure of the wind is not modified, i.e., it maintains the same mass-loss rate, but the super-critical part is accelerated. Comparing the mass-loss rate, we obtained for the first model $\dot{M} \simeq 1.2 \times 10^{-7}$~M$_\odot$~yr$^{-1}$, while for the second $\dot{M} \simeq 1.3 \times 10^{-7}$~M$_\odot$~yr$^{-1}$.

\section{Conclusions}\label{sec.conc}

We presented here some characteristics of Alfv\'en waves and showed how these waves can be
used as an acceleration mechanism of stellar winds.
As a particular case, we evaluated the wind temperature and velocity profiles
of a typical K5 supergiant star using an outward-directed flux of Alfv{\'e}n waves as the
main acceleration mechanism of the wind. For a typical K5 supergiant star, we obtained
a high mass-loss rate and a low terminal velocity consistent with
observations.

It is important to notice that after the recent report on
the unambiguously detection of Alfv\'en waves in the solar corona, these waves
will increase their applicability in the Sun and in several astrophysical environments.

\section*{Acknowledgments}

AAV and VJP thank the Brazilian agencies FAPESP (under grant 04/13846-6)
and CNPq (under grant 304523/90-9) for financial support, respectively. 

\section*{References}

\par\noindent
\hangindent=0.5 true cm
Alfv\'en, H. 1942, Nature, 150, 405.
\par\noindent
\hangindent=0.5 true cm
Balogh, A., Smith, E. J., Tsurutani, B. T., Southwood, D. J., Forsyth,
R. J., \& Horbury, T. S. 1995, Science, 268, 1007.
\par\noindent
\hangindent=0.5 true cm
Banerjee, D., Teriaca, L., Doyle, J. G., \& Wilhelm, K. 1998, A\&A, 339, 208.
\par\noindent
\hangindent=0.5 true cm
Carpenter, K. G., Robinson, R. D., \& Judge, R. G. 1995, ApJ, 444, 424.
\par\noindent
\hangindent=0.5 true cm
Carpenter, K. G., Robinson, R. D., Harper, G. M., Bennett, P. D., Brown, A., \& Mullan, D. J. 1999, ApJ, 521,382.
\par\noindent
\hangindent=0.5 true cm
De Pontieu, B., McIntosh, S. W., Carlsson, M., Hansteen, V. H., Tarbell, T. D., Schrijver, C. J., Title, 
A. M., Shine, R. A., Tsuneta, S., Katsukawa, Y., Ichimoto, K., Suematsu, Y., T. Shimizu, T., \& S. Nagata, S.
2007, Science, 318, 1574.
\par\noindent
\hangindent=0.5 true cm
dos Santos, L. C., Jatenco-Pereira, V., \& Opher, R. 1993, ApJ, 410, 732.
\par\noindent
\hangindent=0.5 true cm
Eaton, J. A. 1993, ApJ, 404, 305.
\par\noindent
\hangindent=0.5 true cm
Esser, R., Lie-Svendsen, \O., Janse, {\AA}. M., \& Killie, M. A. 2005, ApJ, 629, 61.
\par\noindent
\hangindent=0.5 true cm
Falceta-Gon\c calves, D., Vidotto, A., \& Jatenco-Pereira, V. 2006, MNRAS, 368, 1145.
\par\noindent
\hangindent=0.5 true cm
Hartmann, L., \& MacGregor, K. B. 1980, ApJ, 242, 260.
\par\noindent
\hangindent=0.5 true cm
Hartmann, L., Edwards, S., \& Avrett, E. 1982, ApJ, 261, 279.
\par\noindent
\hangindent=0.5 true cm
Hartmann, L., \& Avrett, E. H. 1984, ApJ, 284, 238.
\par\noindent
\hangindent=0.5 true cm
Hollweg, J. V. 1973, ApJ, 181, 547.
\par\noindent
\hangindent=0.5 true cm
Hollweg, J. V. 1987, ApJ, 312, 880.
\par\noindent
\hangindent=0.5 true cm
Holzer, T. E., Fla, T., \& Leer, E. 1983, ApJ, 275, 808.
\par\noindent
\hangindent=0.5 true cm
Jatenco-Pereira, V., \& Opher, R. 1989a, A\&A, 209, 327.
\par\noindent
\hangindent=0.5 true cm
Jatenco-Pereira, V., \& Opher, R. 1989b, MNRAS, 236, 1.
\par\noindent
\hangindent=0.5 true cm
Jatenco-Pereira, V., \& Opher, R. 1989c, ApJ, 344, 513.
\par\noindent
\hangindent=0.5 true cm
Jatenco-Pereira, V., Opher, R., \& Yamamoto, L. C. 1994, ApJ, 432, 409.
\par\noindent
\hangindent=0.5 true cm
Kuin, N. P. M., \& Hearn, A. G. 1982, A\&A, 114, 303.
\par\noindent
\hangindent=0.5 true cm
Lamers, H. J. G. L. M., \& Cassinelli, J. P. 1999, Introduction to
stellar winds (New York: Cambridge University Press.
\par\noindent
\hangindent=0.5 true cm
Lie-Svendsen, \O., Hansteen, V. H., \& Leer, E. 2002, ApJ, 566, 562.
\par\noindent
\hangindent=0.5 true cm
Linsky, J. L. 1998 in ESA Special Publication, Vol. 413, Ultraviolet Astrophysics Beyond the IUE Final Archive, ed. W. Wamsteker, R. Gonzalez Riestra \& B. Harris, 83
\par\noindent
\hangindent=0.5 true cm
Morton, D. C. 1967, ApJ, 147, 1017.
\par\noindent
\hangindent=0.5 true cm
Morton, D. A., Jenkins, E. B. \& Brooks, N. 1969, ApJ, 155, 875.
\par\noindent
\hangindent=0.5 true cm
Okamoto, T. J., Tsuneta, S., Berger, T. E., Ichimoto, K., Katsukawa, Y., Lites, B. W., Nagata, S., K. Shibata, K., 
Shimizu, T., Shine, R. A., Suematsu, Y.,  T. D. Tarbell, T. D. \& Title, A. M. 2007, Science, 318, 1577.
\par\noindent
\hangindent=0.5 true cm
Parker, E. N. 1960, ApJ, 132, 821.
\par\noindent
\hangindent=0.5 true cm
Peter, H., \& Judge, P. G. 1999, ApJ, 522, 1148.
\par\noindent
\hangindent=0.5 true cm
Roddier, F., \& Roddier, C. 1985, ApJ, 295, 21.
\par\noindent
\hangindent=0.5 true cm
Schmutzler, T., \& Tscharnuter, W. M. 1993, A\&A, 273,318.
\par\noindent
\hangindent=0.5 true cm
Smith, E. J., Balogh, A., Neugebauer, M., McComas, D. 1995, Geophys. Res. Lett., 22, 3381.
\par\noindent
\hangindent=0.5 true cm
Tomczyk, S., McIntosh, S. W., Keil, S. L., Judge, P. G., Schad, T., D. H. Seeley, D. H., \& Edmondson, J. 2007,
Science, 317, 1192.
\par\noindent
\hangindent=0.5 true cm
Tu. C.-Y., Shou, C., Marsch, E., Xia, L.-D., Zhao, L., Wang, J.-X., \& Wilhelm, K. 2005, Science, 308, 519.
\par\noindent
\hangindent=0.5 true cm
Underhill, A. B. 1983, ApJ, 268, L127.
\par\noindent
\hangindent=0.5 true cm
Vidotto, A., \& Jatenco-Pereira, V. 2006, ApJ, 639, 416.
\end{document}